\documentstyle[12pt]{article}
\author{Hans-J\"urgen Schmidt }
\title{An alternate Hamiltonian formulation of fourth--order
 theories and its application to cosmology}
\date{}
\begin{document}
\maketitle

\medskip

\centerline{
 Universit\"at Potsdam, Institut f\"ur Mathematik,
 Projektgruppe  Kosmologie}
\centerline{
      D-14415 POTSDAM, PF 601553, Am Neuen Palais 10, Germany}
\centerline{e-mail \  \  \   hjschmi@rz.uni-potsdam.de }

\medskip

\begin{abstract}
An alternate Hamiltonian $H$ different from Ostrogradski's one
 is found  for the Lagrangian $L=L(q, \dot q, \ddot q)$,
where $\partial \sp 2 L / \partial (\ddot q)\sp 2 \ne 0 $. We
 add a suitable divergence to $L$ and insert $a=q$ an
d
 $b=\ddot q$. Contrary to other approaches no constraint is
 needed because $\ddot a = b$ is one of the canonical
 equations. Another canonical equation becomes equivalent to
 the fourth--order Euler--Lagrange equation of $L$.
Usually, $H$ becomes quadratic in the momenta, whereas the
 Ostrogradski approach has Hamiltonians always linear in the
 momenta.

      For non--linear $L=F(R)$, $G=dF/dR \ne 0$ the
 Lagrangians $L$ and $\hat L = \hat F ( \hat R)$ with
$\hat F  =  2R/G\sp 3  -  3 L/G \sp 4 $,
 $\hat g_{ij} = G \sp 2 \, g_{ij}$ and
$\hat R =  3R/G\sp 2  -  4 L/G \sp 3 $
give conformally equivalent fourth--order field equations
 being dual to each  other. This generalizes Buchdahl's result
 for $L=R^2$.

     The exact fourth--order gravity cosmological solutions
 found by Accioly and Chimento are interpreted from the
 viewpoint of the instability of fourth--order theories and
how they transform under this duality.

     Fi\-nal\-ly, the al\-ter\-nate Ha\-mil\-to\-nian is
ap\-plied to de\-duce the Whee\-ler--De Witt equa\-tion for
fourth--or\-der gra\-vi\-ty mo\-dels more
syste\-ma\-ti\-cal\-ly than be\-fore.
\end{abstract}

This is Preprint Uni Potsdam Inst. f. Mathematik Nr. 94/12
in the updated version from 17. 1. 95

of discrete systems: general mathematical aspects

\bigskip

\section{Introduction}

\medskip

Higher--order theories, especially fourth--order gravity
theories, are subject to conflicting facts: On the one hand,
they appear quite naturally from generally accepted
principles; on the other hand, they are unstable and so, they
should be considered unphysical.

\medskip

     Here, we want to attack this conflict from two
directions. First: The Ostrogradski approach [1] to find a
Hamiltonian formulation for a higher--order theory is the most
famous (see e.g. refs. [1 - 9]) but possibly not the best
method. To check this hypothesis, we present an alternate
Hamiltonian formalism for fourth--order theories in sct. 2. It
systematizes what has been sporadically done in the literature
for special examples.

\medskip

     Sct. 3 deals with fourth--order gravity following from a
non--linear Lagrangian $L(R)$. The conformal equivalence of
these theories to theories of other types is widely known, but
the conformal equivalence of these theories to theories of the
same type but essentially different Lagrangian is much less
known. We fill this gap by proving a duality theorem between
pairs of such fourth--order theories in subsection 3.1. The
instability of these theories from the point of view of the
Cauchy problem is subject of subsection 3.2.

\medskip

Sct. 2 applies to arbitrary theories, sct. 3 to gravity, and
both are applied to fourth--order cosmology in sct. 4. We
re--interpret known exact solutions (Friedmann models in
subsection 4.1 and Kantowski--Sachs models in 4.2) under the
stability criteria mentioned before.

\medskip

In the final sct. 5 we discuss quantum effects and give hints
(which shall be outlined in a future paper with S. Reuter) how
to apply the alternate Hamiltonian formalism of sct. 2 to the
Wheeler--De Witt equation for a cosmological minisuperspace
model within fourth--order gravity.

\bigskip

     \ The rest of this intro\-duction short\-ly re\-views
pa\-pers on higher--or\-der theo\-ries.  Eliezer and Woodard
[1] and  Jaen, Llosa and Molina [2]
represent  standard papers for the generalization of the
Ostrogradski approach to non--local systems (see also [3]) and
to systems with constraints (see also [4 - 6]) applying
Dirac's approach.

      Let the Lagrangian $L$ be  a function of the vector
$q_{\alpha}$ and its first $n$ temporal derivatives
$\dot q_{\alpha}$, $\ddot q_{\alpha}$,
$\dots , \,  q^{(n)}_{\alpha}$.
The Hessian is
\begin{equation}
H_{\alpha \beta}=\frac{\partial^2L}{\partial q_{\alpha}
^{(n)}\partial q_{\beta}^{(n)}}
\end{equation}
and the non--vanishing of its determinant
defines the regularity of $L$. In the following we do not
write the subscript $\alpha$; one can think of $q$ as being a
point particle in a (one-- or higher--dimensional) space. In
the Ostrogradski approach, $Q = \dot q$ is taken as additional
position variable. This leads to an ambivalence of the
procedure, because it is not trivial to see at which places
$\dot q$ has to be replaced with $Q$, cf. [7]. We prevent this
ambivalence in our alternate Hamiltonian, cf. sct. 2, by
putting $Q=\ddot q$.

Ref. [8] discusses higher--order field theories. The problem
is the lack of an energy bound, typically two kinds of
oscillators with different signs of energy exist. Usually, one
restricts the space of initial conditions to prevent negative
energy solutions. The authors of ref. [8]  redefine the energy
analogous to the Timoshenko model, so one gets a positive
mechanical energy inspite of an indefinite Ostrogradski
Hamiltonian, they write: "An appealing aspect of this approach
is the absence of any constraint." So it has this property in
common with our approach sct. 2, but it is otherwise a
different one.

     \ \ A second standard procedure [2, 8, 9] for deal\-ing
with higher--or\-der Lagran\-gians is to con\-si\-der them as
a se\-quence in a para\-meter $\epsilon$, so one can break the
Euler--Lagrange--equa\-tion into a se\-quence of se\-cond
or\-der ones. In [9] this is called "reduction of
higher--order Lagrangians by a formal power series in an
ordering parameter." [9] deals also with the Lie--K\"onigs
theorem: a local Hamiltonian is always possible,  and they
consider  some global questions.

     \ \ Let us repeat the famous counter--example [10]
Douglas (1941): it is an example of a second order system not
following from a Lagrangian:
$$\ddot x + \dot y =0 \quad \quad \ddot y + y + \epsilon \dot
x =0$$
It follows from the Lagrangian
$$L=\frac{1}{2} [\dot y^2 - y^2 + \epsilon(
x \dot y - \dot x y - \dot x^2 )]$$
for $\epsilon \ne 0$ and has no Lagrangian otherwise. We
mention this example to show that the following recipe need to
to work always. Recipe for higher--order theories: "Write down
the Euler--Lagrange equations, break them into a sequence of
second order ones by introducing further coordinates. Find
Lagrangians for these second order equations."

\medskip

A powerful method for dealing with  a classical Lagrangian
\begin{equation}
L=\frac{1}{2} \, g_{ij} \, \dot q^i \, \dot q^j - V(q)
\end{equation}
is given in [11] and shall be applied in cosmological
minisuperspace models like in sct. 5. The Euler--Lagrange
equation  to Lagrangian (1.2) reads
$$
\ddot q^i + \Gamma ^i _{jk} \dot q^j \dot q^k
= - g^{ik} V _{,k}
$$
and is fulfilled for geodesics in the Jacobi--metric
$$ \hat g_{ij} \ = \ (E-V) \, g_{ij}$$
Remark: For constant potentials $V$ this is trivial, for
non-constant potentials the constant $E$ must be correctly
chosen to get the result, for $E=V$ it breaks down, of course.

\medskip

  \ \  Stelle [12] cites Ostrogradski [1] but uses other
methods to extract different spin modes for fourth--order
gravity. In [13], a regular reduction of fourth--order gravity
similar to the method with an ordering parameter mentioned
above has been proposed as follows:
In the Newtonian limit one has $$\Delta \Phi + \beta \Delta
\Delta \Phi = 4\pi G \rho ,$$
then one restricts to solutions which can be expanded into
powers of the coupling parameter $\beta $. Argument: If $\beta
$ is a parameter, this is well justified, if it is a universal
constant, then this restriction is less satisfying.
Comment: This restriction excludes the usual Yukawa--like
potential $\frac{1}{r} \exp (-r/\sqrt \beta )$, so one may
doubt whether this method gives the right solutions. Let us
further mention ref. [14] for non--local gravitational
Lagrangians like $L=R\Box^{-1}R + \Lambda$ in two dimensions
and refs. [15, 16] for the linearized $R^2$--theory.

\bigskip

To facilitate the reading of sct. 2, we pick up the example
 eq. (5) of [5]:
\begin{equation}
\tilde L = [ \ddot q \sp 2 + 4 \ddot q \dot q \sp 2 + 4 \dot q
\sp 4 ] e \sp{3q}
\end{equation}
The equation of motion is [5, eq. (6)]
\begin{equation}
2 q \sp{(4)} + 12 \dot q q \sp{(3)} + 9 ( \ddot q )\sp 2 + 18
\dot q \sp 2 \ddot q = 0
\end{equation}
A good check of the validity of the formalism is the
following: For a constant $c>0$ and $\dot q>0$, each solution
of
$$\ddot q = - 2 \dot q \sp 2 + c \sqrt{ \dot q }$$
is also a solution of eq. (1.4).

 By adding a divergence to eq. (1.3) one gets $L = (\ddot
q)\sp 2 e\sp {3q}$. The alternate formalism requires to use
$q\sp 1 = q $ and $q \sp 2 = \ddot q $ as new coordinates. So
we get
\begin{equation}
L = (q\sp 2)\sp 2 \ \exp (3 q\sp 1)
\end{equation}
Eq. (1.5) represents the ultralocal Lagrangian mentioned in
[5]. It is correctly stated in [5], that the alternate
formalism  does not work for this version  eq. (1.5) of the
system. This  clarifies that the addition of a divergence to a
higher--order Lagrangian  sometimes influences the
applicability of the  alternate Hamiltonian formalism.
So one should add a "suitable" total derivative to the
Lagangian. "Suitable"  means, that the space of
solutions is the same at both sides, and that the relation
between the various coordinates is ensured without imposing
any constraints. It turns out, that the Lagrangian $\hat L$
differing from $\tilde L$, eq. (1.3) by a divergence only
\begin{equation}
\hat  L = - [ \ddot q \sp 2 + 6 \ddot q \dot q \sp 2 + 2 \dot
q q\sp {(3)} ] e \sp{3q}
\end{equation}
does the job. Of course, the variations of $L$, $\tilde L$,
and $\hat L$ with respect to $q$ all give the same equation of
motion (1.4). But only in version (1.6) the alternate
formalism (insertion of the equation $Q=\ddot q$ and then
apply the usual formulas of classical mechanics - to avoid
ambiguities with the square--sign we have replaced $q^1$ by
$q$ and $q^2$ by $Q$) leads correctly to the Hamiltonian [5,
eq. (7)]:
\begin{equation}
H=-\frac{1}{2}(pP - 3 P^2 Q)e^{-3q} + Q^2 e^{3q}
\end{equation}
It essentially differs from the Ostrogradski approach because
terms only linear in the momenta do not appear; so one of the
criteria for unboundedness of energy fails to be fulfilled.
The integrability condition $Q=\ddot q$ and the equation of
motion (1.4) both follow from the canonical equations of eq.
(1.7); no constraint is necessary to get this.

\bigskip

\section{The alternate Hamiltonian formalism}
\setcounter{equation}{0}

\bigskip

Let us consider the Lagrangian
\begin{equation}
L \ = \ L(q, \dot q , \ddot q)
\end{equation}
for a point particle $q(t)$, a dot denoting $\frac{d}{dt}$ and
$$q\sp{(n)} =  \frac{d\sp n q}{dt\sp n}$$
The corresponding Euler--Lagrange  equation reads
\begin{equation}
0 = \frac{\partial L}{\partial q} - \frac{d}{dt}
\frac{\partial L}{\partial \dot q} +   \frac{d\sp 2}{dt \sp 2}
\frac{\partial L}{\partial \ddot q}
\end{equation}
We suppose this Lagrangian to be non-degenerated, i.e., $L$ is
non-linear in $\ddot q $. The highest-order term of eq. (2.2)
is
$$q\sp{(4)} \frac{\partial \sp 2 L}{\partial (\ddot q) \sp
2}$$
therefore, non-degeneracy (= regularity, cf. eq. (1.1)) is
equivalent to require that eq. (2.2) is of fourth order, i.e.
$$ \frac{\partial \sp 2 L}{\partial (\ddot q) \sp 2} \ne 0$$
(If $q$ is a vector consisting of $m$ real components then
this condition is to be written as Hessian determinant.)

 If we add the divergence
$\frac{d}{dt} G(q, \dot q)$ to $L$, we do not alter the
Euler--Lagrange equation (2.2). Furthermore, the expression
$\frac{d}{dt} G$ is linear in $\ddot q$, and so its addition
to $L$ does not influence the condition of non--degeneracy.
The addition of such a divergence can therefore simply
absorbed by a suitable redefinition of $L$.

     In the next two subsections we add a special and a more
general divergence to get a Hamiltonian formulation different
from Ostrogradski's one. In the preprint by Kasper [4] a
similar consideration has been made at the Lagrangian's
level. Subsection 2.1 represents only a special case of
subsection 2.2, but we write it down, because it has the
advantage that the formulas can be given explicitly, and so
the formalism becomes more transparent.

\subsection{A special divergence}
     The addition of the following divergence is no more done
by a redefinition of $L$
\begin{equation}
L_{div} = \frac{d}{dt}[ f(q) \, \dot q \, \ddot q], \ \ f(q)
\ne 0
\end{equation}
and we consider $\hat L = L + L_{div}$. The Euler--Lagrange
equation is again eq. (2.2). Using
$$f'(q) \equiv \frac{df}{dq}$$
we get
\begin{equation}
\hat L = L + f'(q)\dot q \sp 2 \ddot q +
f(q)[(\ddot q)\sp 2 + \dot q q \sp{(3)} ]
\end{equation}
which contains third derivatives of $q$.

     We introduce new coordinates
\begin{equation}
a \ = \ q\, , \ b \ = \ \ddot q
\end{equation}
(In the Ostrogradski approach, the second coordinate is $\dot
q$, instead.)
It is obvious that there is exactly this one compatibility
condition:
\begin{equation}
\ddot a \ = \ b
\end{equation}
Let us insert eq. (2.5) into eq. (2.4). This insertion becomes
unique by the additional requirement that $\hat L $ does not
depend on second and higher derivatives of $a$ and $b$, i.e.,
$$\hat L \ = \  \hat L (a, \dot a, b, \dot b ) $$
giving
\begin{equation}
\hat L = L(a, \dot a, b) + f'(a) \dot a \sp 2 b +
f(a)[b\sp 2 + \dot a \dot b ]
\end{equation}
(In the Ostrogradski approach, there remains an ambivalence
which of the $\dot q$ in the original Lagrangian is to be
interpreted as second coordinate and which as time derivative
of the first one.)

     The momenta are defined as in classical mechanics by
\begin{equation}
p_a = \frac{\partial \hat L}{\partial \dot a}, \
p_b = \frac{\partial \hat L}{\partial \dot b}
\end{equation}
(In the Ostrogradski approach, an additional term is
necessary.)
Inserting eq. (2.7) into eqs. (2.8) we get
\begin{equation}
p_a = \frac{\partial  L}{\partial \dot a}
+ 2 f'(a) \dot a b + f(a)\dot b
\end{equation}
and
\begin{equation}
p_b = f(a) \dot a
\end{equation}
Because of $f(a) \ne 0 $, cf. eq. (2.3), we can invert eq.
(2.10) to
\begin{equation}
\dot a = \frac{p_b}{f(a)}
\end{equation}
Inserting eq. (2.11) into eq. (2.9) and dividing by $f(a)$ we
get
\begin{equation}
\dot b = \frac{1}{f(a)}[p_a - \frac{\partial  L}{\partial \dot
a}
- 2 f'(a) b \frac{p_b}{f(a)}]
\end{equation}
It is instructive to make a more general consideration: The
question, whether eqs. (2.9, 10) can be inverted to $\dot a$,
$\dot b$, can be answered by calculating the Jacobian
\begin{equation}
J = \frac{\partial (p_a, p_b)}{\partial (\dot a, \dot b)}
= \frac{\partial p_a}{\partial \dot a}
\frac{\partial p_b}{\partial \dot b}
- \frac{\partial p_a}{\partial \dot b}
\frac{\partial p_b}{\partial \dot a}
\end{equation}
We insert eqs. (2.9, 10) into eq. (2.13) and get
\begin{equation}
J \ = \ - \ [f(a)]\sp 2
\end{equation}
Because of $f \ne 0$ one has also $J \ne 0$ and the inversion
is possible. This more general consideration gave the
additional information that the Jacobian is always negative;
this may be the hint to some kind of instability.

     We define the Hamiltonian $H$ as usual by
$$H \ = \ \dot a p_a + \dot b p_b - \hat L$$
i.e., with eq. (2.7) we get
\begin{equation}
H  =  \dot a p_a + \dot b p_b -  L - f'(a) \dot a \sp 2 b -
 f(a) [ b\sp 2 + \dot a \dot b ]
\end{equation}

Here we insert $\dot a$ according to eq. (2.11) and get the
Hamiltonian $H \ = \ H(a, p_a, b, p_b)$.
The factor of  $\, \dot b$ in $H$ automatically vanishes, so
we do not need eq. (2.12).
The canonical equations read
\begin{equation}
\frac{\partial H}{\partial p_a} \ = \ \dot a
\end{equation}
further
\begin{equation}
\frac{\partial H}{\partial p_b} \ = \ \dot b
\end{equation}
and
\begin{equation}
\frac{\partial H}{\partial a} \ = \ - \dot p_a
\end{equation}
and
\begin{equation}
\frac{\partial H}{\partial b} \ = \ - \dot p_b
\end{equation}

The whole procedure is intended to give the following results:
The Hamiltonian $H$ shall be considered to be a usual
Hamiltonian for two interacting point particles $a(t)$ and
$b(t)$. One of the canonical equations shall be equivalent to
the compatibility condition eq. (2.6) and another one shall be
equivalent to the original Euler--Lagrange equation (2.2),
whereas the two remaining canonical equations are used to
eliminate the momenta $p_a$ and $p_b$ from the system.
     The next step is to find those Lagrangians $L$ which make
this procedure work.
{}From eqs. (2.15) and (2.11) we get
\begin{equation}
H=\frac{p_a p_b}{f(a)} - L(a,\frac{p_b}{f(a)}, b) -
\frac{p_b \sp 2 f'(a)b}{f(a)\sp 2} - f(a)b\sp 2
\end{equation}
In this form, eq. (2.16) coincides with eq. (2.11) and (2.17)
with (2.12). So we may use eqs. (2.9, 10) in the following,
because they are equivalent to eqs. (2.11, 12).

Now, we use eqs. (2.19), cancel  $p_b$ by use of eq. (2.10)
and get  \begin{equation}
0 = \frac{\partial L}{\partial b} + 2 b f(a) - \ddot a f(a)
\end{equation}
In order that the compatibility relation eq. (2.6) follows
automatically from eq. (2.21), one has to ensure that $f(a)
\ne 0$ (which is already assumed) and that
$$0 = \frac{\partial L}{\partial b} +  b f(a)$$
identically takes place. The condition of non--degeneracy,
$$ \frac{\partial \sp 2 L}{\partial b \sp 2} \ne 0$$
 is then also automatically fulfilled. One has the following
possible Lagrangian
\begin{equation}
L = - \frac{1}{2} f(a) b\sp 2 + K(a, \dot a )
\end{equation}
where $K$ is an arbitrary function, but, for simplicity, we
put $K=0$.

The last of the four canonical equations to be used is eq.
(2.18)
reading now with eqs. (2.9, 10, 20)
\begin{equation}
0 = f \ddot b + 2 f' \ \dot a \dot b
+ \frac{3}{2} f' \ b\sp 2 + f'' \ \dot a \sp 2 b
\end{equation}
If we insert here eq. (2.5) we get exactly the same as the
Euler--Lagrange equation (2.2) following from the Lagrangian
\begin{equation}
L = - \frac{1}{2} f(q) (\ddot q)\sp 2
\end{equation}

Result: For every Lagran\-gian of type (2.1) which can be
brought in\-to type (2.24) with $f\ne 0$ the addi\-tion of the
diver\-gence (2.3) makes it pos\-sible to apply the new
coor\-di\-na\-tes (2.5). Then the sys\-tem be\-comes
equi\-valent to a clas\-si\-cal Ha\-mil\-tonian of two
particles, and the relation (2.6) between them follows without
imposing an additional constraint.

\bigskip

\subsection{A general divergence}

In this subsection we try to generalize the result of the
previous subsection by avoiding to prescribe the special
structure (2.3) of the divergence to be added. We substitute
eq. (2.3) by
\begin{equation}
L_{div} = \frac{d}{dt} h(q, \, \dot q , \, \ddot q)
\end{equation}
Keeping eqs. (2.5) we get instead of eq. (2.7) now
\begin{equation}
\hat L = L(a, \dot a, b) + h_1 \dot a  +
h_2 b + h_3 \dot b
\end{equation}
where $h_n$ denotes the partial derivative of $h$ with respect
to its $n$th argument. Using eqs. (2.8), (2.10) is now
replaced with
\begin{equation}
p_b = h_3 (a, \dot a, b)
\end{equation}
Eq. (2.13) is kept, and (2.14) is replaced with
\begin{equation}
J = - (h_{23})\sp 2
\end{equation}
We have to require that $h_{23} \ne 0$, and then the equation
$p_b = h_3$ is locally invertible as
$\dot a  = F(p_b, a, b)$. From this definition one immediately
gets the identity $F_1 \, h_{23} = 1$. Two further identities
to be used later are not so trivial to guess. To derive them,
let us for a moment fix $p_b$ and then calculate the increase
of $h_3$ with increasing $a$ and $b$ resp. The assumed
constancy of $h_3$  yields the equations
\begin{equation}
h_{13} \ + \ F_2 \, h_{23} \ = \ 0
\end{equation}
and
\begin{equation}
h_{33} \ + \ F_3 \, h_{23} \ = \ 0
\end{equation}
resp.
to be used for deducing the generalization of eq. (2.21). One
gets the result: For $h_{23} \ne 0$ (which is already
presumed), the compatibility relation (2.6) follows
automatically from the canonical
equation (2.19) if and only if
\begin{equation}
0 = L_3 + h_2
\end{equation}
is identically fulfilled. One can see: The condition of
 non--degeneracy of the Lagrangian (2.1)  namely
$$L_{33} \ne 0$$
 is equivalent to the condition $h_{23} \ne 0$. For any given
non--degenerate Lagrangian we can find the appropriate
divergence by solving eq. (2.31) as follows
\begin{equation}
h(q, \dot q, \ddot q) = - \int _0 \sp {\dot q}
L_3(q, x, \ddot q) dx
\end{equation}
All other things are fully analogous:
\begin{equation}
H=[p_a - h_1(a, F, b)]F - h_2(a, F, b)b - L(a, F, b)
\end{equation}
where $F=F(p_b, a, b)$.
Eq. (2.19) with (2.30) gives the compatibility condition
(2.6). Eq. (2.18) with (2.29) is equivalent to the
Euler--Lagrange  equation (2.2).

\medskip

     \ \  Let us summarize this section: For the Lagrangian
$L=L(q, \dot q, \ddot q)$ where $\partial \sp 2 L / \partial
(\ddot q)\sp 2 \ne 0 $ we define
 $\hat L = L + L_{div}$ where
$$L_{div} \ = \  - \, \frac{d}{dt} \int \frac{\partial
L}{\partial \ddot q} (q, x, \ddot q) dx $$
We insert $a=q$ and $b=\ddot q$, define the momenta
$p_a = \frac{\partial \hat L}{\partial \dot a}$ and
$p_b = \frac{\partial \hat L}{\partial \dot b}$ and get the
Hamiltonian $H = \dot a p_a + \dot b p_b - \hat L$. One of its
canonical  equations is $\ddot a = b$ and another one is
equivalent to the fourth--order Euler--Lagrange equation
following  from $L$. By these properties, $L_{div}$ is
uniquely determined up to the integration constant. Contrary
to other approaches, no constraint is needed.

\bigskip

\section{Fourth--order gravity}

In Rainich (1925, ref. [17]) the electromagnetic field was
calculated from the curvature tensor. This was
cited in Kucha\v r (1963, ref. [18]) as example for the
geometrization programme; in [18]
on meson fields $\psi$ (now called scalar fields), Kucha\v r
gives a kind of geometrization by using a relation between
$\psi$ and $R$, then he gets the equation
$$\Box R - \frac{k^2}{2} R=0$$
which is of fourth order in the metric. It looks like
fourth--order gravity as we are dealt with, but he does not
deduce it from a curvature squared action.

\subsection{Duality theorems}
\setcounter{equation}{0}

In  Bekenstein  (1974, ref. [19])
the conformal transformation from Einstein's theory with a
minimally coupled ($\phi $) to a conformally coupled ($\psi$)
scalar field is proven where additional conformally invariant

matter (radiation) is allowed. For $8\pi G=1$ one has
$$\psi \ = \ \sqrt 6 \tanh ( \phi /\sqrt 6 )$$
 If radiation is absent then it works also with "coth" instead
of "tanh". This is reformulated in his theorem 2:
If $g_{ij}$ and $\psi$ form an Einstein--conformal scalar
solution, then $\hat g_{ij} = \frac{1}{6} \psi^2  g_{ij}$ and
$\hat \psi = 6/ \psi$ form a second one. One can see that this
is a dual map because by applying the operator $\ \hat{} \ $
twice, the original solution is re--obtained.

Let us comment this theorem 2: For the conformal scalar field
one has the effective gravitational constant $G_{eff}$ defined
by
$$ \frac{1}{8 \pi G_{eff}} \ = \ 1 \  -  \  \frac{\psi^2}{6}$$
A positive value $G_{eff}$ implies a negative value
$\hat G_{eff}$. By changing the overall sign of the Lagrangian
one can achieve a positive effective gravitational  constant
at the price of the scalar field becoming a ghost (wrong sign
in front of the kinetic term). So, Bekenstein has given a
conformal transformation from Einstein's theory with a
conformally coupled ordinary scalar field to Einstein's theory

with a conformally coupled ghost. From this property one can
see that this duality relation is different from the duality
theorem to be deduced at the end of this subsection, because
there one has effectively ordinary scalar fields at both
sides.

Later but independently of [19] the
conformal equivalence between minimally and conformally
coupled scalar fields with $G_{eff} > 0$ was generalized in
[20] by the inclusion of several self--interaction terms.

The conformal transformation from fourth--order gravity to
Einstein's theory with a minimally coupled scalar field was
deduced in several steps:  Bicknell (1974, ref.[21]) found the
  transformation for $L=R^2$; the conformal factor is $R$, and
after the transformation one gets Einstein's theory with
non--vanishing $\Lambda$-term and a massless minimally coupled
scalar field as source.
 Next steps see e.g. [22]. In [23] besides the  conformal
transformation it is shown that the  trace $\sim R^2$ leads to
a term like $R^2\ln R$ in $L$. Ref. [24] generalizes it

  by the inclusion of non--minimally coupled scalar fields.
Jakubiec and Kijowski (1988/89, ref. [25])
generalize the conformal transformations to more general
 transformations of the metric.

  Buchdahl (1978, ref. [26]) showed: For $L=R^2$ the conformal
factor $R^2$ (if it is $\ne 0$) transforms solutions to
solutions and represents a dual map in the set of solutions.
(Another conformal factor than in [21]!) [27] generalizes this
dual map to other non--linear Lagrangians $L(R)$, the
conformal factor being $(\frac{dL}{dR})^2$.
(Again, this conformal factor is the square of that  conformal
factor which is necessary to transform to Einstein's theory
with a minimally coupled scalar field.)

A further type of transformations was presented by
 Buch\-dahl in 1959, ref. [28].
For a space-time $V_n$ pos\-ses\-sing a non-null
hy\-per\-sur\-face--or\-tho\-go\-nal Kil\-ling vec\-tor one
can pro\-duce so\-lu\-tions with a sca\-lar field as follows.
Let the Killing vector be
$\frac{\partial}{\partial x^1}$, the metric is
$g_{kl}$ with $g_{kl,1}=0$, $g_{11} \ne 0$ and
$g_{1\alpha} =0$ where greek indices take all values except
$1$. Let $R_{kl} =0$ and the dimension $n\ge 2$ but $n \ne 3$.
We fix two reals $A$ and $B$ and define a new metric
$\tilde g_{kl}$ by $\tilde g_{1\alpha} =0$,
$$\tilde g_{11} = (g_{11})^A \, \quad  \ \tilde g_{\alpha
\beta} = (g_{11})^B g_{\alpha \beta}$$
(We consider the case $g_{11} > 0$, the other sign is treated
analogously.) This is a conformal transformation for $A=B+1$.
Defining $g_{11} = e^{2\psi}$ one has $R_{11}=0$ iff (= if and
only if)
$\Box \psi =0$. So, $\Box \psi =0$ is presumed from the
beginning.

Further, it holds: $\tilde{} \ \Box \psi =0$ is then
identically fulfilled iff $$A=1-B(n-3)$$
(For $n=4$, this means $A=1-B$.)

$B=0, \ A=1$ represents the identical transformation. For
non--identical transformations the equation is compatible with
a conformal transformation for $n=2$ only. This is in
agreement with the fact that the D'Alembert operator is
conformally invariant for $n=2$ only.

Buchdahl's results are: For $n=4$ and $A=-1$, i.e., $B=2$, one
gets a vacuum solution which is in general different from the
initial one. For $n=4$ and $\vert A \vert \ne 1$, one gets a
solution of Einstein's theory with a massless minimally
coupled scalar field (which is proportional to $\psi$) as
source. This supplements [21], because here no $\Lambda$-term
is needed.

\medskip

In [29, 30] the problem is discussed which of the conformally
equivalent metrics in these theorems is the physical metric.
Magnano and  Soko\l owski  (1994, ref. [30]) represents a
good review to this theme.

\bigskip

Let us now deduce the duality theorem announced in the
introduction which shall close a gap in the set of  the

aforementioned results.

\medskip

Let
\begin{equation}
\hat g_{ij} = e \sp {2U} \ g_{ij}
\end{equation}
and
$\Box _c \ = \ \Box - \frac{R}{6}$. Then
\begin{equation}
\hat{} \, \Box _c \ = \ e \sp {-3U} \ \Box _c \ e \sp {U}
\end{equation}
reflects the conformal invariance of the operator $\Box _c $
if applied to a scalar. The following consideration is
restricted to the case that both $R$ and $\hat R $ are
positive; the other sign can be dealt analogously.

\medskip

We put $U=\ln R$ into eq. (3.1) and apply the identity (3.2)
to the constant scalar $=1$. We multiply by $(-6R)$ and get
the identity
\begin{equation}
R \ \hat R \ = \ 1 \ - \ \frac{6}{R^2} \, \Box R
\end{equation}
Further, we define the operator  \ $\hat{}$ \ given by
\begin{equation}
\hat g_{ij} = R^2 \ g_{ij}
\end{equation}
to be a dual one if it coincides with its inverse operator.
Then the following conditions are equivalent:

1. \ $\hat{}$ \ is a dual operator.

2. $R \ \hat R \ = \ 1$

3. $\Box R = 0$

4. $ \hat{} \, \Box  \hat R = 0$

\medskip

{\bf Proof}. 1. $ \Leftrightarrow $ 2. is simply the explicit
form of the definition of duality. 2.
  $ \Leftrightarrow $ 3. follows from eq. (3.3). 3.  $
\Leftrightarrow $ 4. is a consequence of the duality property.

\medskip

{\bf Remarks}. 1. The conformal factor $R^2$ in eq. (3.4) is
crucial for the validity of the duality. Already for spaces
with constant curvature scalar one can see:
$\hat R = \frac{1}{R}$ requires the conformal factor to be
$R^2$. \  2. This consideration was inspired by Buchdahl's
 paper [26] from 1978.

\medskip

Now we are prepared to show a duality relation between pairs
of fourth--order gravity theories. (We formulate it only for
dimension $4$, other dimensions $>2$ give similar results.)
 Let $L=F(R)$ with
$G=\frac{dF}{dR} \ne 0$, $H=\frac{dG}{dR} \ne 0$ and
$\hat g_{ij} = G^2 \ g_{ij}$. If  $g_{ij}$ is a solution of
the fourth--order equation following from $L$ then it holds
\begin{equation}
\hat R \ = \ 3R/G\sp 2 \ - \ 4 F/G \sp 3
\end{equation}
One has $\frac{d\hat R}{dR} \ne 0$ iff (= if and only if)
\begin{equation}
G^2 \ \ne \ 6H(2F - GR)
\end{equation}
{\bf Remark}. If $R$ is a constant then $2F=GR$ follows from
the field equation, (3.6) is automatically fulfilled, and
(3.5) reduces to the known relation $\hat R = R/G^2$.

Let (3.6) be fulfilled in the following. Then eq. (3.5) can be
inverted locally as $R=R(\hat R)$. We define
\begin{equation}
\hat F \ = \ 2R/G\sp 3 \ - \ 3 F/G \sp 4
\end{equation}
where $R=R(\hat R)$ has to be inserted into the r.h.s.

\medskip

The following theorem holds under the presumptions formulated
above.

\medskip

{\bf Theorem. $\hat L \, = \, \hat F(\hat R)$ defines a
fourth--order theory dual to $L=F(R)$ and $\hat g_{ij}$ is a
solution of its field equation.}

\medskip

 Duality means that by applying this procedure twice the
original theory with original solution $ g_{ij}$ is obtained.

\medskip

{\bf Remarks}. 1. The most problematic step in formulating
this theorem was to find the formulas (3.5) and (3.7); the
existence of such a theorem was announced in [27], but these
two crucial formulas are presented here for the first time.

 2. We call two theories $F, \, \tilde F$ to be similar (i.e.,
they go into each other by a change of the length unit) if
there exist non--vanishing reals $\alpha$ and $\beta$ such
that
$\tilde F(R)=\alpha F(\beta R)$. It holds: Similar theories
 have similar duals.

\medskip

{\bf Proof of the Theorem}. The trace of the field equation
following from $L$ reads $$GR \ + 3 \Box G \ = \ 2 F$$ It must
be supplemented by the condition that the trace--free part of
the tensor $GR_{ij} - G_{;ij}$ vanishes. We use identities
analogous to eq. (3.3) with $G$ instead of $R$ and the
relation holding for the r.h.sides of eqs. (3.5) and (3.7):
$$\frac{d\hat R}{dR} \ = \ G \,   \frac{d\hat F}{dR}$$
 This ensures the validity of
$$\hat G \, = \,
\frac{d\hat F}{d \hat R} \, = \, \frac{1}{G} \, \ne \, 0$$
The rest of the proof is lengthy but straightforward.

\bigskip

{\bf Examples}. 1. Let $L=\frac{R^m}{m}$ with $m\ne 0, \, 1$.
Unequality (3.6) excludes $m=\frac{3}{2}$ and
$m=\frac{4}{3}$. For all remaining reals $m$ one gets the dual
$\hat L = c \hat R ^{\hat m}$ with
$\hat m \ = \ \frac{3m-4}{2m-3}$  and a constant $c=c(m)$, cf.
[27].

\medskip

2. Let $L=R\, - \, \alpha R^2$. The inversion of eq. (3.5) is
not possible in closed form, but in the vicinity of flat space
we may expand into powers of $\hat R$ and get up to third
order
$$\hat L = \hat R \, - \, \alpha \hat R^2 - 4 \alpha ^2
\hat R^3$$
Result: At least a special type of cubic terms in the
Lagrangian can be absorbed by a suitable conformal
tranformation.

\bigskip

\subsection{Instability of $R^2$-theories}

This sub\-section deals with the clas\-si\-cal
insta\-bi\-li\-ty of fourth--order theo\-ries fol\-lowing from
a non--line\-ar Lagran\-gian $L(R)$.

(Quan\-tum in\-sta\-bi\-lities will be commented in sct. 5.)

Teys\-san\-dier and Tour\-renc (1983, ref. [31]) sol\-ved the
Cau\-chy--prob\-lem for this theo\-ry, let us short\-ly
re\-peat the main in\-gre\-dients.

The Cauchy problem is well--posed (a property which is usually
required to take place for a physically sensible theory) in
each interval of $R$-values where both $dL/dR$ and $d^2 L /
dR^2$ are different from zero. The constraint equations are
similar as in General Relativity: the four $0i$-component
equations. What is different are the necessary initial data to
 make the dynamics unique. More exactly: Besides the data of
General Relativity one has to prescribe the values of $R$ and
$\frac{dR}{dt}$ at the initial hypersurface. This coincides
whith the general experience: Initial data have to be
prescribed till the highest--but--one temporal derivative
appearing in the field equation (here: fourth--order field
equation,   $\frac{dR}{dt}$ contains third--order temporal
derivatives of the metric). Under this point of view,
classical stability of the field equation means that a small
change of the Cauchy data implies also a small change of the
solution.

Now we are prepared to classify the stability claims found in
refs. [32 - 36]. To simplify we specialize to the Lagrangian
$L=R - \epsilon R^2$ with the non--tachyonic sign
$\epsilon > 0$ and restrict to the range $\frac{dL}{dR} > 0$,
i.e. $R<\frac{1}{2\epsilon}$

\medskip

On the one hand, refs. [32, 33]  find a classical instability
of the Minkowski space--time for this case. (Mazzitelli and
Rodrigues [33] cite  Gross,  Perry and Yaffe (1982, ref. [34])
with the sentence "The Minkowski solution in general
relativity has been proven to be stable." which refers to the
positive energy theorem of general relativity.)

\medskip

On the other hand, refs. [35, 36] find out that the  Minkowski
space--time is here not more unstable than in General
Relativity itself. What looks like a contradiction from the
first glance is only a notational ambivalence as can be seen
now: The main argument in refs. [32, 33] is that an
arbitrarily  large value $\frac{dR}{dt}$ is compatible with
small values of $H^2$ and $R^2$. In refs. [35, 36] however,
following the Cauchy--data argument [31], ($\frac{dR}{dt}$
being part of the Cauchy data which are presumed to be small)
stability of the Minkowski space--time is obtained in the
version: If the Cauchy data are small (meaning: close to the
Cauchy data of the Minkowski space--time) then the
fourth--order field equation bounds the solution to remain
close to the Minkowski space--time.

The argument of ref. [35] is a little bit different: There the
conformal transformation to Einstein's theory with a scalar
field $\Phi$ [22] is applied; it is observed that in the
$F(R)$-theory there are never ghosts which implies stability.
Now, $\Phi$ and $\frac{d\Phi}{dt}$ belong to the Cauchy data
which is equivalent to the data $R$, $\frac{dR}{dt}$ in the
conformal picture thus supporting the Cauchy data argument
given at the beginning of this subsection.

\bigskip

\section{Cosmology}
\setcounter{equation}{0}

Several papers [37 - 43] apply the conformal transformation
theorem [21, 22] to cosmology; so for interpreting the
cosmological singularity [37], for dealing with
 anisotropic models [38], with transformation to Brans--Dicke
extended inflation [40].
Ref. [42] mentions that $\Omega_0 < 1$ is possible even if
$k=1$. The other ones apply the theorem mainly as a
mathematical device to transform solutions to solutions of the
other theory.

In the following two subsections we discuss exact solutions
directly found for fourth--order gravity (subsection 4.1: a
spatially flat Friedmann model, subsection 4.2: a
Kantowski--Sachs model).

\bigskip

\subsection{Friedmann models}

In 1988, Chimento [44] found an exact solution for
fourth--order gravity in a spatially flat Friedmann model. He
also found out that in the tachyonic--free case the asymptotic
matter-dominated Friedmann solution is stable, and no
fine--tuning of initial conditions is necessary to get the
final (oscillating) Friedmann stage; particle production of
non--conformal fields may backreact to damp the oscillations.

[45] generalizes [44]: here the Dirac equation is considered,
the result is that there appear spinor field oscillations and
the qualitative behaviour remains essentially the same.

Let us present the exact solution of [44]. For the spatially
flat Friedmann model with Hubble parameter
$H=\dot a/a$ he solves the fourth--order field equation with
vacuum polarization term.
The zero--zero component equation reads
\begin{equation}
2H\ddot H - \dot H^2 + 6 H^2\dot H + \frac{9}{4} H^4 + H^2=0
\end{equation}
The $H^4$-term stems from the vacuum polarization and the
$H^2$-term from the Einstein tensor. The remaining ingredients
of eq. (4.1) come from the term $R^2$ in the Lagrangian. (Here
we only present the tachyonic--free case with $\Lambda =0$ and
$\frac{9}{4}$ in front of $H^4$.) The factor in front of $H^4$
should not influence the weak--field behaviour because for
$H\approx 0$ this factor only changes the effective
gravitational constant.

{}From eq. (4.1) the discussion of subsection 3.2 becomes
obvious: (4.1) represents a third--order equation for the
cosmic scale factor $a$; it is a constraint and not a
dynamical equation. (It is only due to the high symmetry, that
accidentally the validity of the constraint implies the
validity of the dynamical equation.) Supposed, eq. (4.1) would
be  the true dynamical equation for a theory, then the
instability argument of [32] could apply.

 The ansatz for solving eq. (4.1)
$$H=\frac{2\dot s}{3s}$$
leads to a non--linear third--order equation for $s$
\begin{equation}
2 \dot s s^{(3)} - \ddot s ^2 + \dot s^2 =0
\end{equation}
Derivative with respect to $t$ yields
the  equation $s^{(4)} + \ddot s=0$
being linear in $s$ and having the solution
$$s= c_1 + c_2 t + c_3 \sin (t + c_4)$$
Inserting this solution into the original equation gives the
restriction $\vert c_2\vert = \vert c_3\vert$.
Let us discuss this solution:
$c_2=0$ leads to the uninteresting flat space--time.
So, now let $c_2 \ne 0$. Adding $\pi$ to $c_4$ can be absorbed
by a change of the sign of $c_3$. Therefore, $c_2=c_3$ without
loss of generality. Multiplication of $s$ by a constant factor
does not change the geometry, so let $c_2=1$. A suitable
time--translation leads to $c_1=0$. Finally, the cosmic scale
factor is calculated as $a=s^{2/3}$ leading to
\begin{equation}
a\ = \ [ t +  \sin (t + c_4) ]
^{2/3} \ \sim \ t^{2/3} \, [1 + \frac{2}{3t} \sin (t + c_4) ]
\end{equation}
The r.h.s. of eq. (4.3) gives in an elegant way the late--time
behaviour already deduced in [46]. The factor $1/t$ in front
of the  "sin"-term shows that the oscillations due to the
higher--order terms are damped. The total energy "sitting" in
these oscillations, however, remains constant in time (because
of the volume--expansion) cf. Suen (1994, ref. [32]) and can
be converted into classical matter by particle creation.

\medskip

Let us mention some further cosmological solutions with
higher--order gravity: [47] discusses the $L(R)$-stability
with a conformally coupled scalar field. Ref.
[48] (partial results of it can be found in [49]) deals with
fourth--order cosmological models of Bianchi--type I and
power--law metrics, i.e.
$$ds^2 = dt^2 - \sum_{i=1} ^3 \ t^{2p_i} \ (dx^i)^2 $$
with real parameters $p_i$. The suitable notation
$$a_k =  \sum_{i=1} ^3 \ p_i ^k$$
gives the following: $a_1=a_2=1$ is the usual Kasner solution
for Einstein's theory. $a_1^2 + a_2 = 2 a_1$ is the condition
to be fulfilled for a solution in $L=R^2$. Refs. [50, 51] also

 discuss  $R^2$-models. Br\"uning, Coule and  Xu (1994, ref.
[36]) consider inflationary cosmology with a Lagrangian
$$ L \ = \  R  + \ \lambda R_{\mu \nu} R^{\mu \nu}/R$$
and mention that it is unclear under which circumstances the
existence of the Weyl term in anisotropic models allows the de
Sitter space--time to be an attractor solution.
Ref. [52] deals with anisotropic Bianchi--type IX solutions
for $L=R^2$. They look for chaotic behaviour analogous to the
mixmaster model in Einstein's theory. Ref. [53] gives exact
solutions for $L=R^2$ and a closed Friedmann model, ref.
[54] discusses the bounce in closed Friedmann models for $L=R
-  \epsilon R^2$. Supplementing the discussion of [54, eq.(1)]
let us mention: In the non--tachyonic case, there exist
 periodically oscillating models with an always positive scale
factor $a$. Ref. [55] looks for chaos in isotropic models,
e.g. by conformally coupled massive scalar fields in the
closed universe. The papers [56, 57] consider the stability of
power--law inflation for $L=R^m$ within the set of spatially
flat Friedmann models. Refs. [58] give overviews on
higher--order cosmology, especially chaotic inflation as an
attractor solution in initial--condition space. [59] deals
with quantum gravitational effects in the de Sitter
space--time, and [60] gives a classification of
in\-fla\-tio\-na\-ry Ein\-stein--sca\-lar--field--mo\-dels via
ca\-ta\-stro\-phe theory. Ref. [61] considers  Chern--Simon
terms in Bianchi cosmologies and the cosmic no-hair
conjecture. The axion with field strength
$H_{ijk}$ puts an extra hair on black holes. Its
energy-momentum tensor does not fulfil the energy conditions,
and so one gets both recollapsing solutions and ever-expanding
solutions which are essentially anisotropic also for late
times.

\medskip

\subsection{Kantowski--Sachs models}

Before we come to the fourth--order solution by Accioly let us
mention some results on Kantowski--Sachs models in general.

The solution found in 1950, cf. ref. [62] - now it is called
Nariai solution - is the only static spherically symmetric
solution of the Einstein equation with positive $\Lambda$-term
which cannot be written in Schwarz\-schild coordinates. It has
a six-dimensional isometry group and is of Kantowski--Sachs
type; it represents the direct product of two two--dimensional
spaces of equal and non--vanishing constant curvature (in
short: $S^2 \times S^2$). One of the many possibilities to
present it is
\begin{equation}
ds^2 = (1-\Lambda r^2)dt^2 - \frac{dr^2}{1-\Lambda r^2} -
\frac{1}{\Lambda^2}(d\theta^2 + \sin^2\theta d\phi^2)
\end{equation}
Ref. [63] discusses the stability of the  Bertotti--Robinson
(also the direct product of two two--dimen\-sio\-nal spa\-ces
of   non--va\-ni\-shing constant cur\-va\-ture, but with
va\-ni\-shing 4--cur\-va\-ture sca\-lar, in short  $\tilde S^2
\times S^2$)
 and Nariai solutions. Ref. [64] is an obituary to H. Nariai,
it  especially mentions the Nariai solution.  Kofman,  Sahni
and Starobinsky (1983, ref. [65]) get the  result that  there
is no particle production in the Nariai solution. [66]
discusses the analogous question for the Bertotti--Robinson
metric, the authors of ref. [67] consider the Nariai metric
and its decay into de Sitter and Kasner--like space--times.
They consider essentially Einstein's vacuum theory with
positive $\Lambda$--term and Kantowski--Sachs metric.

Torrence and Couch  (1988, ref. [68])
 showed how the de Sitter space-time can be presented locally
as a Kantowski--Sachs cosmological model. Moreover, no other
 Robertson--Walker space--time can be presented as
Kantowski--Sachs model, cf. also [69] for this question.

 Moniz (1993, ref. [70]) discussed the
 Kantowski--Sachs models in Einstein's theory with positive
$\Lambda$--term in relation to  the no--hair conjecture and
got the following result. The majority of solutions is
asymptotically de Sitter, a small number recollapses: infinite
to finite measure if $dB \, d\dot B$ is taken as measure in
the initial condition space, and $B$ is the radius of the
$S^2$ of the model:
$$
ds^2 = dt^2 -  A^2(t) dr^2 - B^2(t) (d\theta^2 + \sin^2\theta
d\phi^2)
$$
  Accioly (1987/88, refs. [71, 72]) found an exact
cosmological solutions of the G\"odel type for higher--order
gravity, the new solution he found is
the direct product of a 3--space of non--vanishing constant
curvature with the real line in space--like direction (see
[72, eq. (18)]). This gives a 7--dimensional isometry group.

Comment: Because of the high symmetry, this exact solution
belongs both to the G\"odel and to the Kantowski--Sachs
classes of metrics (with $A=$ const. and $B=\cosh t$). The
latter class is more popular, so we classify it here and not
primarily as G\"odel model.

 The direct product of a 3--space of po\-si\-tive  con\-stant
cur\-va\-ture  with the real line in time--like di\-rec\-tion
is known as Ein\-stein's sta\-tic uni\-verse. So, by
imaginary coordinate transformations, both metrics can be
mapped onto each other locally. This implies that Accioly's
solution is conformally flat (because Einstein's universe
carries this property). In the set of conformally flat
space--times, the term $R_{ij}R^{ij}$ gives the same
variational derivative as $\frac{1}{3} R^2$, so Accioly's
paper has to be classified in the set of non--linear
Lagrangians $L=F(R)$; he assumes $F$ to be a quadratic
function of $R$. As one knows there is a critical value of the
curvature scalar in such theories, it is defined by
$\frac{dF}{dR}=0$. (Cf. subsection 3.2 above: there the Cauchy
problem is not well--posed.)

 At these values, the fourth--order differential equation
reduces to the second order one $F=0$; it turns out that
Accioly's solution obeys this critical value so it is unstable
similar to Einstein's static universe.
More directly: Let $L=(R-R_0)^2$ with a constant $R_0$, then
each solution of the second--order equation $R=R_0$ solves
also the fourth--order field equation following from $L$.
  So a lot of solutions (including Accioly's one) can be
found, but they all live in the region where the Cauchy
problem is ill--posed.

\medskip

\section{Summary}
\setcounter{equation}{0}

The scope of this paper was to present the foundations
necessary to deduce the Wheeler--de Witt equation for a
cosmological minisuperspace model in fourth--order gravity.

The method (sporadically developed in [51] for $L=R^2$ and a
spatially flat Friedmann model) to handle with eqs. (1.3 -
1.6) was systematically generalized in sct. 2 to give a
Hamiltonian formulation of a general fourth--order theory. The
possibility of deducing this method makes it clear that the
method of ref. [51] is not restricted to highly symmetric
models. The alternate Hamiltonian formulation has some
advantages in comparison with Ostrogradski's one: No
constraint is needed, the Hamiltonian is typically a quadratic
function in the momenta. (Ostrogradski's approach leads always
to a Hamiltonian linear in the momenta which gives artificial
factors $i$ in the Schr\"odinger equation.) The calculation of
the momenta  from the Lagrangian follows the usual equations
(2.8)  whereas the Ostrogradski approach needs some additional
terms. Our approach is less ambiguous, cf. eq. (2.7).

One could pose the question whether both approaches are
equivalent on another level, this is not fully excluded, but
even if it is the case, the approach deduced here is more
directly applicable to fourth--order quantum cosmology.

The fact that the Jacobian eq. (2.28) is always negative
excludes the possibility to get a positive definite Jacobi
metric in eq. (1.2). This is one of the many possibilities to
say what is meant by the phrase "fourth--order theories are
always unstable". The Jacobi metric plays the role of the
conformally transformed superspace--metric used in quantum
cosmology. And here the circle can be closed: In Einstein's
theory (both for Lorentzian and  Euclidean signature of the
underlying manifold) the superspace--metric has Lorentzian
signature and cannot be positive definite. So we get once more
the result of subsection 3.2: Fourth--order gravity contains
some instabilities,  but only those which it has in common
with General Relativity.

To decide the quantum instability of the Minkowski or de
Sitter space--times in fourth--order gravity one must solve
the corresponding Wheeler--de Witt equations
 (Mazzitelli and Rodrigues [33] deduced them for the spatially
flat Friedmann model and the Lagrangian $L=R-\epsilon R^2$)
 and has to interpret them carefully. This has to be done yet.

In [50] it is mentioned that a classical theory with higher
derivatives has instabilities: "At the quantum level, the
difference is even more dramatic. Noncommuting variables in
the lower--derivative theory, such as position and velocities,
become commuting in the higher--derivative theory." Remark of
U. Kasper to this sentence: "The uncertainty relation is
primarily between positions and momenta. If the momentum is
independent of the velocity then commuting position and
velocity need not bother."

The duality theorem deduced in subsection 3.1 is a method
to construct new solutions of fourth--order gravity from known
solutions of a (possibly other) fourth--order theory. It gives
non--trivial results only for solutions with a non--constant
curvature scalar, e.g. the Chimento solution [44] which has
been rewritten in subsection 4.1. Kantowski--Sachs models
(whose quantum cosmology is considered in [73]) have begun to
discuss in subsection 4.2; this shall be completed elsewhere.

\medskip

{\it Acknowledgement}. I thank Dr. U. Kasper and Dr. M. Rainer
for making some clarifying remarks.  Financial support from
the  Wis\-sen\-schaft\-ler--Inte\-gra\-tions--Pro\-gramm under
contract Nr. 015373/E and from the Deutsche
For\-schungs\-gemein\-schaft under Nr. Schm 911/5-2 is
gratefully acknowledged.

\medskip

{\Large {\bf References}}

[1]  M. Ostrogradski, Memoires Academie St. Petersbourg
Series{\bf  VI} vol.  {\bf 4},  385 (1850); D. Eliezer
and R. Woodard, Nucl. Phys. {\bf B 325}, 389 (1989).

[2] X. Jaen, J. Llosa and A. Molina, Phys. Rev. {\bf D 34},
2302 (1986).

[3] J. Llosa and J. Vives, Int. J. Mod. Phys. {\bf D 3}, 211
(1994).

[4] J. Govaerts, Hamiltonian quantization and constrained
dynamics, Leuven Univ. Press 1991;
 J. Govaerts, Phys. Lett. {\bf B 293}, 327 (1992);
 J. Govaerts and M. Rashid, The Hamiltonian formulation of
higher order dynamical systems, Preprint hep-th/9403009, March
1994; U. Kasper, Class. Quantum Grav. {\bf 10}, 869 (1993);
 U. Kasper, On the Hamiltonian formalism for cosmological
models in fourth order gravity theories, Preprint
Universit\"at Potsdam 94/11, October 1994.

[5] G. Kleppe, Ana\-ly\-sis of higher deri\-va\-tive
hamil\-tonian for\-ma\-lism, Pre\-print UAHEP-9407 Univ. of
Alabama 1994. It represents a comment to ref. [51].

[6] C. Battle, J. Gomis, J. Pons and N. Roman--Roy, J. Phys.
{\bf A 21}, 2693 (1988); V. Nesterenko, J. Phys. {\bf A 22},
1673 (1989).

[7] X. Gracia, J. Pons and N. Roman--Roy, J. Math. Phys.
{\bf 32}, 2744 (1991).

[8] A. Chervyakov and V. Nesterenko, Phys. Rev. {\bf D 48},
5811 (1993).

[9] T. Damour and G. Sch\"afer, J. Math. Phys. {\bf 32}, 127
(1991); V. Perlick,  J. Math. Phys. {\bf 33}, 599 (1992).

[10] J. Douglas, Trans. Amer. Math. Soc. (TAMS) {\bf 50}, 71
(1941).

[11] M. Szydlowski and J. Szczesny, Phys. Rev. {\bf D 50}, 819
(1994).

[12] K. Stelle, Gen. Relat. Grav. {\bf 9}, 353 (1978).

[13] L. Bel and H. Sirousse-Zia, Phys. Rev. {\bf D 32}, 3128
(1985).

[14] J. Navarro-Salas, M. Navarro, C. Talavera and V. Aldaya,
Phys. Rev. {\bf D 50}, 901 (1994).

[15] P. Teyssandier, Class. Quant. Grav. {\bf 6}, 219  (1989);
  P. Teyssandier, Annales de Physique,
  Coll.1, Suppl.6, {\bf 14}, 163  (1989);
 P. Teyssandier, Astron. Nachr. {\bf 311}, 209  (1990).

[16] B. Linet and P. Teyssandier, Class. Quant. Grav. {\bf 9},
159  (1992).

[17] G. Rainich, Trans. Amer. Math. Soc. (TAMS) {\bf 27}, 106
(1925).

[18] K. Kucha\v r, Czech. J. Phys. {\bf B 13}, 551 (1963).

[19] J. Bekenstein, Ann. Phys. NY {\bf 82}, 535 (1974).

[20] S. Deser, Phys. Lett. {\bf B 134}, 419 (1984); H.-J.
Schmidt, Phys. Lett. {\bf B 214}, 519 (1988).

[21] G. Bicknell, J. Phys. {\bf A 7}, 1061 (1974).

[22] B. Whitt, Phys. Lett. {\bf B 145}, 176 (1984);
 G. Magnano, M. Ferraris and M. Francaviglia, Gen. Relat.
Grav. {\bf 19}, 465 (1987);  H.-J. Schmidt, Astron. Nachr.
{\bf 308}, 183 (1987); J. Barrow and S. Cotsakis, Phys. Lett.
 {\bf B 214}, 515 (1988).

[23] M. Baibosunov, V. Gurovich and U. Imanaliev,
 J. eksp. i teor. fiz. {\bf 98}, 1138  (1990).

[24] K. Maeda, Phys. Rev. {\bf D 39}, 3159 (1989); L.
Amendola, M. Litterio and F. Occhionero, Int. J.
Mod. Phys. {\bf A 5}, 3861  (1990).

[25] A. Jakubiec and J. Kijowski, Phys. Rev. {\bf D 37}, 1406
(1988); A. Jakubiec and J. Kijowski, J. Math. Phys. {\bf 30},
1073  (1989); A. Jakubiec and J. Kijowski, J. Math. Phys. {\bf
30}, 2923  (1989).

[26] H. Buchdahl, Int. J. theor. Phys. {\bf 17}, 149 (1978).

[27] H.-J. Schmidt,  Class. Quant. Grav. {\bf 6}, 557  (1989).

[28] H. Buchdahl, Phys. Rev. {\bf 115}, 1325 (1959).

[29] J. Audretsch, A. Economou and C. Lousto, Phys. Rev.
 {\bf D 47}, 3303 (1993); H.-J. Schmidt, Phys. Rev.
 {\bf D 51} in print.

[30] G. Magnano and L. Soko\l owski, Phys. Rev. {\bf D 50},
5039 (1994).

[31] P. Teyssandier and Ph. Tourrenc, J. Math. Phys. {\bf 24},
2793 (1983); A.A. Starobinsky and H.-J. Schmidt, Class. Quant.
Grav. {\bf 4}, 695  (1987).

[32] W.-M. Suen, Phys. Rev. {\bf D 40}, 315 (1989); W.-M.
Suen, Phys. Rev. {\bf D 50}, 5453 (1994); D. Coule and M.
Madsen, Phys. Lett. {\bf B 226}, 31 (1989).

[33] F. Mazzitelli and L. Rodrigues, Phys. Lett. {\bf B 251},
45 (1990).

[34] D. Gross, M. Perry, L. Yaffe, Phys. Rev. {\bf D 25}, 330
(1982).

[35] G. Lopez Cardoso and B. Ovrut, Int. J. Mod. Phys. {\bf D
3}, 215 (1994).

[36]  V. M\"uller, Int. J. Mod. Phys. {\bf D 3}, 241 (1994);
 E. Br\"uning, D. Coule and C. Xu, Gen. Relat. Grav. {\bf 26},
1197 (1994); H.-J. Schmidt, Phys. Rev. {\bf D 50}, 5453
(1994).

[37] G. Le Denmat and H. Sirousse-Zia, Phys. Rev. {\bf D 35},
480 (1987).

[38] L. Pimentel and J. Stein - Schabes, Phys. Lett. {\bf B
216}, 27 (1989).

[39] O. Bertolami, Phys. Lett. {\bf B 234}, 258  (1990).

[40] Y. Wang, Phys. Rev. {\bf D 42}, 2541  (1990).

[41] J. Barrow and K. Maeda, Nucl. Phys. {\bf B 341}, 294
(1990); D. Hochberg, Phys. Lett.  {\bf B 251}, 349  (1990).

[42] D. La, Phys. Rev. {\bf D 44}, 1680 (1991).

[43] U. Bleyer, Proceedings Conference Physical
Interpretations of Relativity Theory, London 1992, p. 32;
 S. Mignemi and D. Wiltshire, Phys. Rev. {\bf D 46}, 1475
(1992);  L. Amendola, D. Bellisai, R. Carullo and F.
Occhionero, p. 127 in:  Relativistic  Astrophysics and
Cosmology,  Eds.:  S. Gottl\"ober, J. M\"ucket, V. M\"uller
WSPC Singapore 1992;  S. Capozziello, L. Amendola and F.
Occhionero, ditto  p. 122.

[44] L. Chimento, Class. Quant. Grav. {\bf 5}, 1137 (1988);
 L. Chimento, Class. Quant. Grav. {\bf 6}, 1285 (1989);
 L. Chimento, Class. Quant. Grav. {\bf  7}, 813 (1990); L.
Chimento, Gen. Relat. Grav. {\bf 25}, 979  (1993).

[45] L. Chimento, A. Jakubi and F. Pensa, Class. Quant. Grav.
{\bf 7}, 1561 (1990).

[46] V. M\"uller and H.-J. Schmidt, Gen. Relat. Grav. {\bf
17}, 769  (1985).

[47] C. Laciana, Gen. Relat. Grav. {\bf 25}, 245
(1993).

[48] H. Caprasse, J. Demaret, K. Gatermann and H. Melenk,
 Int. J. Mod. Phys. {\bf C 2}, 601  (1991).

[49] V. M\"uller, Ann. Phys. (Leipz.) {\bf 43}, 67 (1986).

[50] F. Mazzitelli, Phys. Rev. {\bf D 45}, 2814 (1992).

[51] H.-J. Schmidt, Phys. Rev. {\bf D 49}, 6354 (1994).

[52] P. Spindel and M. Zinque, Int. J. Mod. Phys. {\bf D 2},
279   (1993).

[53] H.-J. Schmidt, Gen.
Relat. Grav. {\bf 25}, 87  (1993), Erratum p. 863.

[54] D. Coule, Class. Quant. Grav. {\bf 10}, L25  (1993).

[55] E. Calzetta, Int. J. Mod. Phys. {\bf D 3}, 167 (1994).

[56] A. Burd and J. Barrow, Nucl. Phys. {\bf B 308}, 929
(1988); C. Sivaram and V. de Sabbata, p. 503 in: Quantum
Mechanics in Curved  Space-Time, Eds. J. Audretsch, V. de
Sabbata, Plenum NY 1990; V.  M\"uller, H.-J. Schmidt and A.A.
Starobinsky, Class. Quant. Grav. {\bf 7}, 1163  (1990);
 Y. Kitada and K. Maeda, Phys. Rev. {\bf D 45}, 1416  (1992);
 J. Aguirregabiria, A. Feinstein and J. Ibanez, Phys. Rev.
{\bf D 48}, 4662 (1993);  J. Aguirregabiria, A. Feinstein and
J. Ibanez, Phys. Rev. {\bf D 48}, 4669 (1993).

[57] S. Cotsakis and P. Saich, Class. Quant. Grav. {\bf  11},
383 (1994).

[58] J. Barrow, Lect. Notes Phys. {\bf 383}, 1  (1991);
 J. Kung and R. Brandenberger, Phys. Rev. {\bf D 42},
1008 (1990).

[59] C. Kiefer, Quantum gravitational effects in de Sitter
 space, Preprint gr-qc/9501001, to appear in:
 New Frontiers in Gravitation, eds. G. Sardanashvily  and R.
 Santilli (Hadronic Press, 1995).

[60] F. Kusmartsev, E. Mielke, Y. Obukhov and
F. Schunck, Preprint gr-qc/9412046, Classification of
inflationary Einstein--scalar--field--models via catastrophe
 theory.

[61] N. Kaloper, Phys. Rev. {\bf D 44}, 2380 (1991).

[62] H. Nariai, Sci. Rep. Tohoku Univ. {\bf 34}, 160 (1950);
  H. Nariai, Sci. Rep. Tohoku Univ. {\bf 35}, 62 (1951).

[63] H. Ishihara and H. Nariai, Progr. Theor. Phys. {\bf 70},
1148 (1983).

[64] K. Tomita, p. 1602 in: Proc. 6th Marcel Grossmann
meeting on General Relativity, Kyoto, Eds.: H. Sato, T.
Nakamura, WSPC Singapore 1992.

[65] L. Kofman, V. Sahni and A. A. Starobinsky, J. eksp. i
teor. fiz. {\bf 85}, 1876 (1983), Sov. Phys. JETP {\bf 58},
1090.

[66] L. Kofman and V. Sahni, Phys. Lett. {\bf B 127}, 197
(1983).

[67] V. Sahni and L. Kofman, Phys. Lett. {\bf A 117}, 275
(1986).

[68] R. Torrence and W. Couch, Gen. Relat. Grav. {\bf 20}, 603
 (1988).

[69] \O . Gr\o n, J. Math. Phys. {\bf 27}, 1490 (1986);
 H. Baofa, Int. J. Theor. Phys. {\bf 30}, 1121 (1991); H.-J.
 Schmidt, Fortschr. Phys. {\bf 41}, 179 (1993).

[70] P. Moniz, Phys. Rev. {\bf D 47}, 4315 (1993).

[71] A. Accioly, N. Cim. {\bf B 100}, 703 (1987).

[72] A. Accioly, Progr. Theor. Phys. {\bf 79}, 1269 (1988).

[73] H.-D. Conradi, Preprint gr-qc/9412049, Trieste.

\end{document}